\documentstyle[12pt,aaspp4,flushrt,psfig]{article}

\newcommand{\asinh}{\sinh^{-1}}
\newcommand{\sgn}[1]{{\hbox{sgn}(#1)}}
\newcommand{\var}[1]{{\hbox{Var}(#1)}}

\begin{document}

\title{%
A Modified Magnitude System that Produces Well-Behaved Magnitudes,
Colors, and Errors Even for Low Signal-to-Noise Ratio Measurements
}

\author{
Robert H. Lupton\altaffilmark{1,2},
James E. Gunn\altaffilmark{2},
and Alexander S. Szalay\altaffilmark{3}}
\altaffiltext{1}{For the SDSS collaboration.}
\altaffiltext{2}{Department of Astrophysical Sciences,
		Princeton University, Princeton, NJ 08544}
\altaffiltext{3}{Department of Physics and Astronomy,
		The Johns Hopkins University, Baltimore MD 21218}

\begin{abstract}
We describe a modification of the usual definition of astronomical
magnitudes, replacing the usual logarithm with an inverse hyperbolic
sine function; we call these modified magnitudes `asinh
magnitudes'. For objects detected at signal-to-noise ratios of greater
than about five, our modified definition is essentially identical to
the traditional one; for fainter objects (including those with a
formally negative flux) our definition is well behaved, tending to a
definite value with finite errors as the flux goes to zero.

This new definition is especially useful when considering the
colors of faint objects, as the difference of two `asinh' magnitudes
measures the usual flux ratio for bright objects, while avoiding
the problems caused by dividing two very uncertain values for
faint objects. 

The Sloan Digital Sky Survey (SDSS) data products will use this scheme
to express all magnitudes in their catalogs.
\end{abstract}

\keywords{catalogs, methods: statistical, techniques: photometric}

\section{Introduction}

The advantages of using a logarithmic scale to measure astronomical
fluxes are obvious: the magnitude scale is able to span a huge dynamic
range, and when relative colors are needed they can be computed by
simply differencing magnitudes measured in different bandpasses. These
advantages are quite clear for bright objects where noise is not an
issue. On the other hand, as fluxes become comparable to the sky
and instrumental noise, the corresponding magnitudes are subject to
large and asymmetric errors due
to the singularity in the magnitude scale at zero flux; indeed, if a
noisy measurement of an object's flux happens to be negative, its
magnitude is a complex number!  Astronomers have generally handled these
cases by specifying detection flags and somehow encoding the negative
flux. These problems become even more pronounced as we work in
multicolor space. An object can be well detected and measured in
several of the bands, and still fail to provide a measurable flux in
others. In flux space the object would be represented as a
multivariate Gaussian probability distribution centered on its measured
fluxes, not necessary
all positive. Important information is lost by simply replacing
a flux by an arbitrary 2- or 3- $\sigma$ upper limit.
In magnitude space, the error
distribution of such an object has an infinite extent in some
directions, making meaningful multicolor searches in a database
impossible; all the non-detections in one or more bands must
be isolated, and treated separately.
One solution to this problem is to use linear units (e.g. Janskys),
although this makes studies based on flux ratios (colors) inconvenient.
This paper proposes an alternative solution, a
modification of the definition of magnitudes, which preserves their
advantages while avoiding their disadvantages.

\section{The Inverse Hyperbolic Sine}

We propose replacing the logarithm in the traditional definition of a
magnitude with an inverse hyperbolic sine function.  This function
becomes a logarithm for large values of its argument, but is linear
close to the origin.
\begin{equation}
    \asinh(x) = \ln \left[ x + \sqrt{x^2+1} \right] \rightarrow 
	\cases{\sgn{x} \ln |2x|, 
		& if $|x|\gg 1$\cr x, 
		& if $|x|\lesssim 1$\cr}
\end{equation}
The usual apparent magnitude $m$ can be written in terms of the
dimensionless normalized flux $x \equiv f/f_0$ as
\begin{equation}
   m \equiv -2.5 \log_{10} x = -(2.5 \log_{10}e) \ln x \equiv -a \ln x.
\end{equation}
where $f_0$ is the flux of an object with magnitude $0.0$, and
$a\equiv 2.5 \log_{10}e = 1.08574$ is Pogson's ratio (Pogson 1856).

Let us define the new magnitude $\mu$ as
\begin{equation}
   \mu(x) \equiv -a \left[ \asinh \left( \frac{x}{2b}\right) + \ln b\right]
\end{equation}
Here $a$ and $b$ are constants, and $b$ is an
arbitrary `softening' which determines the flux
level at which linear behaviour sets in. After some discussion, we have
adopted the name `asinh magnitudes' for $\mu$.
Consider the asymptotic
behaviour of $\mu$, for both high and low $x$:
\begin{equation}
	\lim_{x\to\infty}{\mu(x)} = -a \ln x = m 
		\qquad\qquad\qquad
	\lim_{x\to0}{\mu(x)} = -a \left[{ x\over 2b} + \ln b\right].
\end{equation}

Thus for $x\rightarrow\infty$, $\mu$ approaches $m$, for any choice of
$b$.  On the other hand when $|x| \lesssim b$, $\mu$ is linear in $x$;
for $x \ll -b$, we
gradually return to logarithmic behaviour, although this regime is
never of astronomical interest for any reasonable choice of $b$.
Intuition suggests that $b$ should be chosen to be comparable to the
(normalized) flux of an object with a signal-to-noise ratio of about one;
the following section discusses the choice of $b$,
and the related question of $\mu$'s error distribution.

\section{The Errors in $\mu$ and the choice of the Softening Parameter $b$}
\label{secChooseB}

Although the softening parameter $b$ can be any positive number,
we show below that one particular selection has a couple of attractive
features.  In making our choice, we use the following guiding principles:
1) Since asinh magnitudes are being introduced to avoid the problems
that classical magnitudes manifest in low signal-to-noise data,
the differences between asinh and classical magnitudes should be minimized
for high signal-to-noise data.
2) We should minimize $\mu$'s variance at low flux levels. The latter is not
strictly required as a too-small value of $b$ merely stretches out $\mu$'s
scale along with its errors, but it is convenient if $\mu$'s variance
at zero flux is comparable to its variance at a signal-to-noise ratio
of a few.

In reality, our measurement of the normalized flux $x$ will be noisy,
with variance $\sigma^2$.  We wish to choose $b$ to minimize $\mu$'s
variance, while keeping the difference $m - \mu$ small.
Let us therefore compute the variances of $m$ and $\mu$
(keeping only the linear terms in their Taylor series),
and also their difference;
arrows indicate the asymptotic behavior as $x \rightarrow 0$:
\begin{eqnarray}
\nonumber
	\var{m}=& {\displaystyle \frac{a^2\sigma^2}{x^2}}
		& \rightarrow  \frac{a^2\sigma^2}{x^2} \\
	\var{\mu} =&  {\displaystyle \frac{a^2 \sigma^2}{4b^2 + x^2}}
		&\rightarrow \frac{a^2\sigma^2}{4b^2} \\
\nonumber
	m-\mu\ =&  {\displaystyle
			a \ln \left[\frac{1 + \sqrt{1+4b^2/x^2}}{2}\right]}
		&\rightarrow  -a\, \sgn{x}\ln\left({|x|\over b}\right)
\end{eqnarray}

What are the disadvantages of taking either too low or too high a
value for $b$?  Choosing a low value of $b$ causes the difference
$m-\mu$ to become smaller, i.e. the two magnitudes track each other
better, but unfortunately $\mu$'s variance at $x=0$ varies as
$1/b^2$. Choosing too high a value has the opposite
effect: the difference explodes at low values of $x$, simply due to
the singularity in the logarithm in the definition of $m$. At the same
time, $\mu$'s variance remains small.

In order to balance these two competing effects, we shall determine $b$
by minimizing a penalty function containing terms due to both, added
in quadrature.

\begin{figure}
\psfig{figure=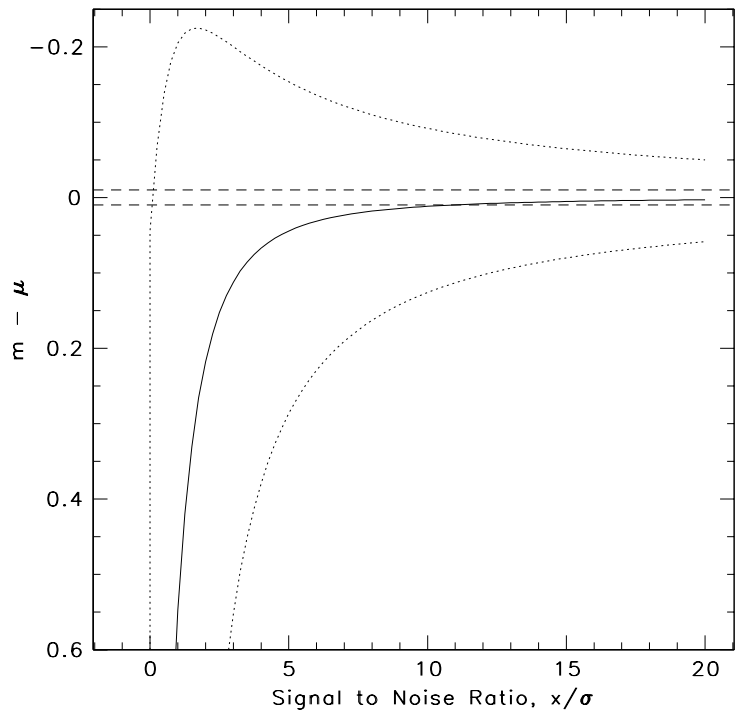,height=4in}
\caption{%
The behavior of $m - \mu$ as a
function of signal-to-noise ratio, $x/\sigma$. The solid line is
the value of $m - \mu$ and the region between the dotted lines
corresponds to the $\pm 1\sigma$ error region for $m$.
The dashed lines are drawn at $\pm 0.01$.
}
\label{figMMu}
\end{figure}

The difference between the two magnitudes, normalised by $m$'s standard
deviation, is given by
\begin{equation}
	\delta(x) = {m(x)-\mu(x)\over \sqrt{\var{m}}} \equiv {b\over\sigma} 
	F\left({x\over b}\right), 
	{\qquad \rm where \qquad}
	F(y) = y \ln \left[ {1 + \sqrt{1+4/y^2}\over 2}\right].
\end{equation}
The function $F(y)$ has a maximum value of approximately 0.5 at $y=0.7624$,
so the largest possible deviation between the two magnitude scales is
\begin{equation}
  \left(m - \mu\right)_{\hbox{max}} \approx \frac{b}{2\sigma}\sqrt{\var{m}}
\end{equation}
The other `cost' associated with the choice of $b$ is the size of the
error box for $\mu$ at $x=0$, which is 
\begin{equation}
	\sqrt{\left.\var{\mu}\right|_{x=0}} = {a \sigma\over 2b}
\end{equation}
The total penalty can be obtained by adding these two costs in quadrature:
\begin{equation}
	\epsilon = \delta_{\hbox{max}}^2 + \left.\var{\mu}\right|_{x=0} =
	{b^2\over 4\sigma^2}+ {a^2 \sigma^2\over 4b^2} =
	{a\over 4} \left[ {b^2\over a\sigma^2} + {a\sigma^2\over b^2}\right]
\end{equation}
which has the obvious minimum at $b^2= a \sigma^2$. Thus the optimal
setting is the value $b= \sqrt{a}\sigma = 1.042\sigma$.
As expected, $b$ is approximately equal to the noise in the flux.
This choice of $b$ leads
to $m - \mu$ having a maximum value of 0.52$\sqrt\var{m}$, implying that
the difference between the two magnitudes is always smaller than the
uncertainty in $m$ (see figure \ref{figMMu}).
If the error in its measured flux is $1\sigma$, the error in $\mu$
is $\pm 0.52$. If the flux errors are Gaussian,
so, to leading order, are the errors in $\mu$ as the transformation
from counts to $\mu$ is linear for $|x| \lesssim b$.

\begin{figure}
\psfig{figure=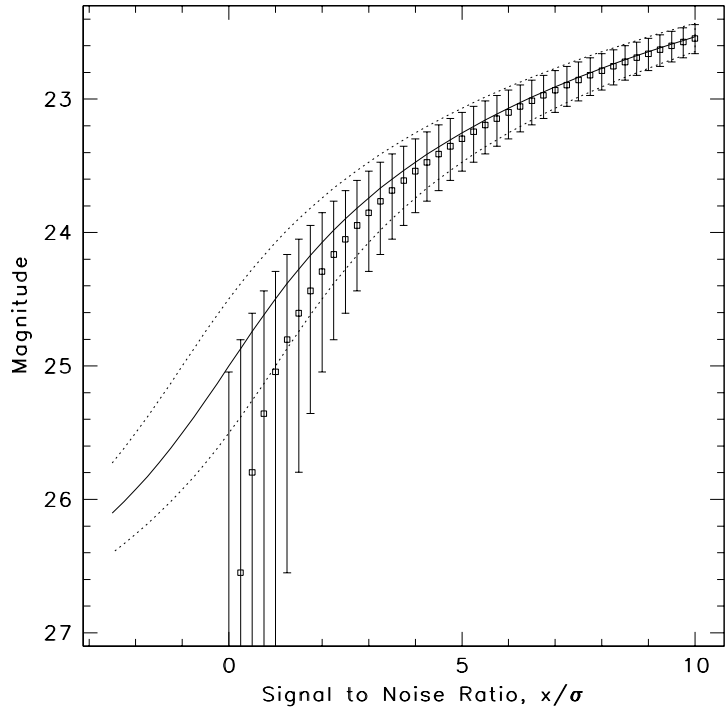,height=4in}
\caption{%
The behavior of $m$ and $\mu$, and their respective errors, as a
function of signal-to-noise ratio, $x/\sigma$. The solid line is
the value of $\mu$ and the
region between the dotted lines its $\pm 1\sigma$ error region;
the points with errorbars are the classical magnitudes, $m$.
We have arbitrarily chosen a zeropoint of $\mu =
25.0$ for an object with no flux.  One other feature of our modified
magnitudes is apparent from this figure, namely that the error band on
$\mu$ is nearly symmetrical, while the errors in $m$ are strongly
skewed at faint magnitudes.
For $S/N$ ratios of less than about two, $m - \mu$ exceeds the value
$0.52\sqrt\var{m}$ quoted in the main body of the paper; this is due to the
breakdown of the linear approximation used to calculate $m$'s variance.
}
\label{figErrors}
\end{figure}

Figure \ref{figErrors} shows $m$ and $\mu$ as a function of signal-to-noise
ratio, for this choice of $b$, along with their 1-$\sigma$ errors.

\section{Application to Real Data}

We have been working in terms of $x \equiv f/f_0$, but it is usually
more convenient to use the measured fluxes directly;
In terms of the {\em non-}normalised flux $f$, the expressions for
$m$, $\mu$, and $\var{\mu}$ become
\begin{eqnarray}
   m =& m_0 - 2.5 \log_{10}f, \\
\label{eqMuZeroPoint}
   \mu =& \left(m_0 - 2.5 \log_{10} b'\right) - a \asinh \left( f/2b' \right)\\
\noalign{and}
\label{muError}
   \var{\mu}\ =& \frac{a^2 \sigma'^2}{4b'^2 + f^2} 
		\approx \frac{a^2\sigma'^2}{4b'^2}
\end{eqnarray}
where
$m_0 \equiv 2.5\log_{10}(f_0)$, $b' \equiv f_0 b$, and
$\sigma' \equiv f_0 \sigma$ are measured in real flux units (e.g. counts).

An object with no measured flux in a given band
has a $\mu$ value of $m_0 - 2.5 \log_{10} b'$ (equation \ref{eqMuZeroPoint}),
in other words the classical magnitude of an object with a flux of $b'$.
We note in passing that this value $\mu(0)$ is a convenient measure of
the depth of a survey, containing information about both the noise properties
of the sky and the image quality.

In the discussion above we considered the idealized case of all
objects having the same error, dominated by sky noise. This case covers
most objects found in a given deep survey, as most are detected at the flux
limit and are typically at most marginally resolved. For bright
objects, of course, the difference between $m$ and $\mu$ is entirely
negligible.

The optimal measure of the flux of a faint stellar object is given
by convolving its image with the PSF. If the noise is dominated by the sky and
detector, the variance of the measured flux is independent of the
object's brightness, and is given by the background variance per unit area
multiplied by the effective area of the PSF ($4\pi\alpha^2$ if the PSF
is Gaussian with FWHM $2\sqrt{2\ln2}\,\alpha$). If we decide upon a
typical seeing quality and sky brightness for a given band, this
defines $\sigma$'s nominal value, $\sigma_0$, which sets $b$ once and
for all. Each band has its own value of $b$.

As observing conditions change so do measurement errors, with the result
that the error in $\mu$ for very faint objects is no longer exactly
the 0.52 magnitudes that it would be under canonical conditions.
Whenever a precise error is needed for a given object's $\mu$,
it may be found by converting $\mu$ back to flux, or by
applying equation \ref{muError}; for faint objects this reduces to multiplying
the quoted error by $\sigma/\sigma_0$. Failure to apply such a correction
would mean that the quoted errors on $\mu$ were wrong.

It would be possible to choose $b$ separately for different parts of
the sky, but this would make the conversion of $\mu$ back to flux
impractically complicated, and a significant source of mistakes for
users of the data. The behaviour of $\mu$ as $b$ changes is
reasonably benign; the error at zero flux varies only linearily
with $b$, as does the flux where $m$ and $\mu$ begin to diverge.

Care is also required whenever the measured flux has
different noise properties, for example if the flux is measured within
a circular aperture or a given isophote rather than using a PSF.
In this case, the appropriate value of $\sigma$ 
may be much larger from the one used to set $b$, with the consequence
that the error in $\mu$ at zero flux considerably underestimates the
true uncertainty (the other case, where the effective aperture is
smaller than the PSF, is unlikely to occur in practice).  It would,
of course, be possible if confusing to choose a different set of $b$
values for each type of (fixed size) aperture, although it seems
unlikely that this would really be a good idea. As the discussion
in section \ref{secChooseB} showed the consequences
of even a grossly incorrect value of $b$ are not catastrophic; the asinh
magnitudes still reduce to our familiar magnitudes for bright objects,
and are still well defined for negative values.

One place where special care will be needed is in measures of surface
brightness, where $m$ and $\mu$ can depart quite strongly from one
another even at levels where the flux is well determined.  It may prove
desirable to use a different value of $b$ for such measurements;
they are after all never directly compared with total magnitudes.

Fan et al. (1999a) have used the asinh magnitude system to search
for high-z quasars in preliminary SDSS data; examples of
color-color plots employing asinh magnitudes may be found in
Fan et al. (1999b).

\section{Asinh Magnitudes and Colors}

The ratio of two low signal-to-noise ratio measurements
(for example, an object's color)
is statistically badly behaved
(indeed, for Gaussian distributions if the denominator has zero mean,
the ratio follows a Cauchy distribution and accordingly has no mean,
let alone a variance!). What is the behavior of our asinh magnitudes when
used to measure colors?

For objects detected at high signal-to-noise ratio, the difference in
$\mu$ measured in two bands is simply a measure of the relative flux
in the two bands.  For faint objects this is no longer true, although
the difference is well behaved.  A non-detection in two bands has a
well defined `color' ($\mu_1(0) - \mu_2(0)$).
As discussed above, the error in this color is approximately $0.75
\sigma/\sigma_0$ magnitudes, assuming independent errors in the
two bands.  Equivalently, such a non-detection can be represented by an
ellipsoid in multi-color space, centered at the point corresponding to
zero flux in all bands, with principal axes $0.52 \sigma/\sigma_0$ (in
general $\sigma/\sigma_0$ will be different in each band).

As an illustration of the instability of the traditional definition
of color for faint objects, consider two objects that have almost
identical colors but which are near the detection limit of a survey. Their
asinh colors will be very similar, but (due to the singularity of the
logarithm as the flux goes to zero) their classical magnitudes may
differ by an arbitrarily large amount.

%
%

\begin{figure}
\psfig{figure=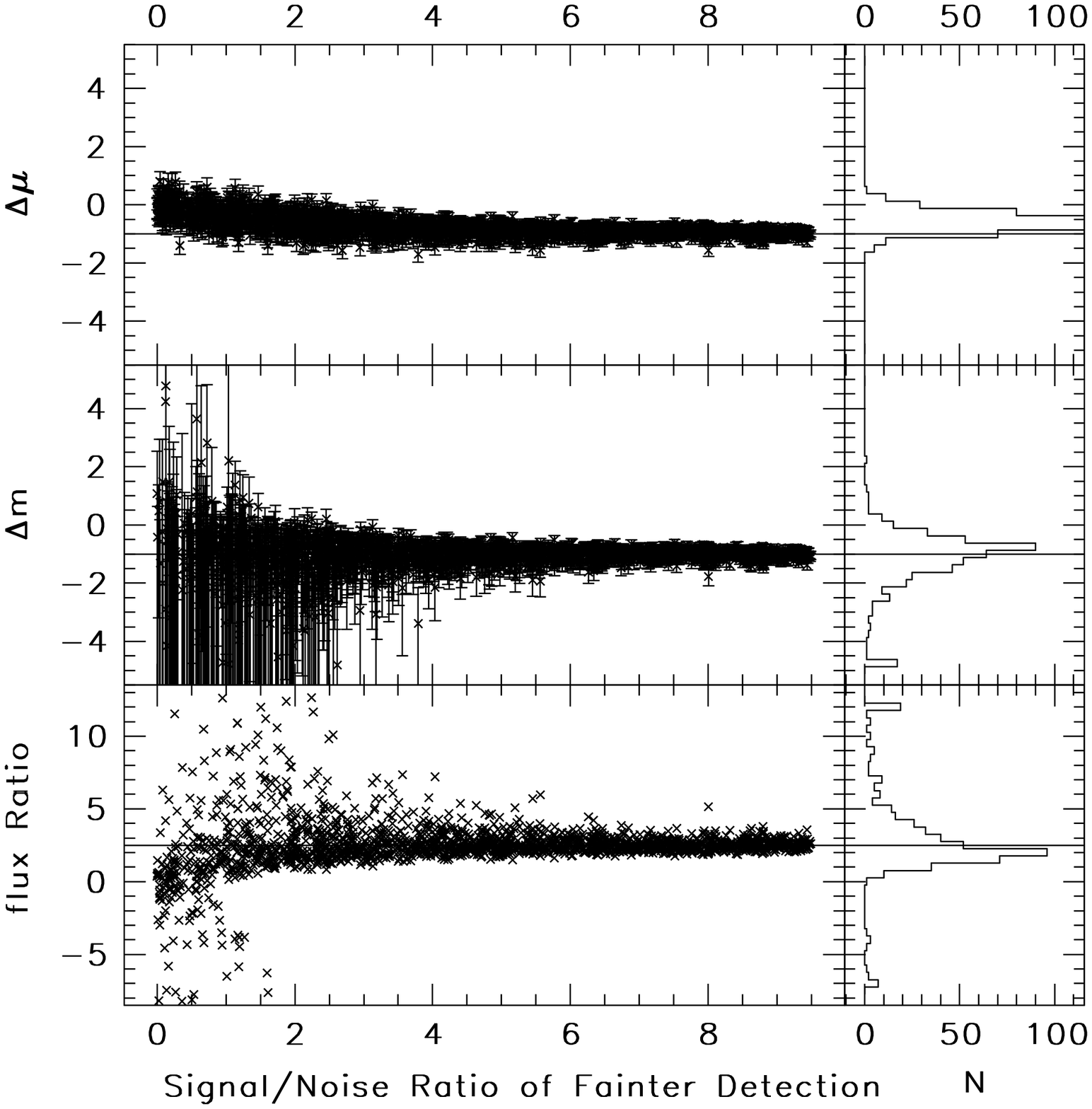,height=4in}
\caption{%
The results of a Monte-Carlo simulation of the colors of a set
of objects.
The bottom panel shows the ratio of the brighter to
the fainter measurement; it is clear that many of the low
signal-to-noise points have negative flux ratios.  The center panel
shows the same points, but now in terms of the magnitude difference of
the two detections. We have simply omitted points for which one or
both fluxes were negative, although they do appear in the histogram.
Finally, the top panel shows the difference
in the asinh magnitudes. The `color' at zero flux was taken to be 0.0
for this simulation.
}
\label{figColor}
\end{figure}

Figure \ref{figColor} shows the results of a simple Monte-Carlo
simulation. We took a set of `objects' with 2.51 times as much flux in
one band in the other (one magnitude), added Gaussian noise of fixed variance
to each measurement, and tabulated the color measured using both classical
and asinh magnitudes. The left hand panel shows the flux ratio, 
$\Delta m \equiv m_1 - m_2$, and $\Delta \mu \equiv \mu_1 - \mu_2$
as a function of signal-to-noise ratio; the right hand panels show
histograms of their distribution in the range $1 <= S/N <= 3$.
At the right side of the plot, where the noise
is less important, both $\Delta m \equiv m_1 - m_2$
and $\Delta \mu \equiv \mu_1 - \mu_2$ tend to -1, the
correct value. As the noise becomes more important, the errors on
the $\Delta m$ plot grow (and an increasing fraction of points in the left
panel is
simply omitted as their fluxes are zero or negative).
The $\Delta\mu$ plot shows the `color' tending to
its value at zero flux, in this case 0.0, as the signal-to-noise ratio
drops.

\section{Summary}

We have shown that an innovative use of inverse hyperbolic
sines for a new magnitude scale can overcome most deficiencies of
traditional magnitudes, while preserving their desirable features.
The defining equations are
\begin{eqnarray*}
   \mu =&
	\left(m_0 - 2.5 \log_{10} b'\right) - a \asinh \left( f/2b' \right) &
	\equiv \mu(0) - a \asinh \left( f/2b' \right) \\
\noalign{and}
   \var{\mu} =& \frac{a^2 \sigma'^2}{4b'^2 + f^2} \approx
			\frac{a^2\sigma'^2}{4b'^2}
\end{eqnarray*}
where $a \equiv 2.5 \log_{10}e$, $f$ is the measured flux, $\sigma'$ the
error in $f$ due to the sky and detector, and $b'$ a softening parameter
(Equations \ref{eqMuZeroPoint} and \ref{muError}). 

The principal advantages of these `asinh' magnitudes are their
equivalence to classical magnitudes when errors are negligible,
their ability to represent formally negative fluxes, and their well behaved
error distribution as the measured flux goes to zero. For high
signal-to-noise ratios the difference of two asinh magnitudes
is a measure of the flux ratio, while for noisy detections it
becomes the statistically preferable flux difference; this allows
meaningful color cuts even when an object is barely detected in some
bands. Additionally, $\mu(0)$ provides a convenient way
of summarizing the photometric depth of a survey for point-like
objects, containing information about both the noise level of the system
and the image quality.

Asinh magnitudes will be used in the SDSS catalogs.

\acknowledgements
The authors would like to thank Xiaohui Fan, Don Schneider, and
Michael Strauss for
helpful comments on various versions of this paper, and Gillian Knapp
for demonstrating that the whole question was more subtle than we had
realised.

The Sloan Digital Sky Survey (SDSS) is a joint project of The University of
Chicago, Fermilab, the Institute for Advanced Study, the Japan
Participation Group, The Johns Hopkins University, Princeton University,
the United States Naval Observatory, and the University of Washington.
Apache Point Observatory, site of the SDSS, is operated by the Astrophysical
Research Consortium.   Funding for the project has been provided by
the Alfred P. Sloan Foundation, the SDSS member institutions,
the National Aeronautics and Space Administration,
the National Science Foundation and the U.S. Department of Energy.

\clearpage

\begin{references}
\reference{} Fan X. et al., 1999a, AJ, submitted.
\reference{} Fan X. et al., 1999b, in preparation.
\reference{} Pogson~N.R., MNRAS {\bf 17}, 12, 1856.
\end{references}
\end{document}